\documentstyle[12pt,bezier]{article}
\begin{document}
\title{Comment on: Cyclic quantum--evolution dependence on
the Hamiltonian and geometric phase}
\author{Ali Mostafazadeh\thanks{E-mail:
alimos@phys.ualberta.ca}\\ \\
Theoretical Physics Institute,
Department of Physics,\\
University of Alberta,
Edmonton, Alberta,\\
Canada T6G 2J1.}
\date{April 1996}
\maketitle

\begin{abstract}
It is shown that the analysis and the main result of the article
by L-A. Wu [Phys.\ Rev.\ A.\ {\bf 53}, 2053 (1996)] are
completely erroneous.
\end{abstract}

PACS number: 03.65.Bz
\vspace{.5cm}
\baselineskip=18pt

There are a number of  mistakes in the article by Wu \cite{wu}.
I shall point out the most obvious ones which question the merits
of the basic idea presented in this article.

According to Eq.~3 of \cite{wu}, it is the phases $e^{i\phi_j}$ which
are single-valued functions, not the phase angles $\phi_j$. The phase
angles are not single-valued. Therefore, Eq.~5 and consequently
Eq.~6 of \cite{wu}, in which $N_j$ are treated as single-valued
integers, are not valid as strict equations. Working with  
single-valued
quantities, i.e., $e^{i\phi_j}$, one can show that Eq.~4 of \cite{wu}
together with the completeness and orthonormality of $\{\psi_j\}$,
lead to:
	$$C_j(e^{i\phi_j}-e^{i\phi})=0\;, ~~~~~{\rm for ~all}  
~~j\;.$$
Therefore, for all $j$ with $C_j\neq 0$, one has $e^{i\phi_j}=
e^{i\phi}$. But this means that:
	\begin{equation}
	\psi(\tau)=e^{i\phi}\sum_jC_j\psi_j=e^{i\phi}\psi(o)\;.
	\label{q1}
	\end{equation}
On the other hand, since $\psi(\tau)$ is the solution of the
Schr\"odinger equation:
	\begin{equation}
	\psi(\tau)=U(\tau,0)\psi(0)\;.
	\label{q2}
	\end{equation}
Eqs.~(\ref{q1}) and (\ref{q2}) indicate that contrary to the
claim made in Ref.~\cite{wu}, $\psi(0)$ is an eigenstate
vector of $U(\tau,0)$. In fact, using the completeness of
$\{\psi_j\}$, one can conclude that $\psi(0)$ must be
proportional to one of $\psi_j$'s, i.e., $C_j=0$ for all $j$
except one. The only exception to this argument is the case
where $e^{i\phi}$ is a degenerate eigenvalue of $U(\tau,0)$.
In this case, there may be more than one $C_j$ that is
non-vanishing. But then the geometric phase is a matrix
belonging to the unitary group $U(n)$, where $n$ is the
degree of degeneracy, and again the analysis of
Ref.~\cite{wu} does not apply.

Finally I would like to comment on a remark made in Ref.~\cite{wu}
(first paragraph) regarding the Berry's phase being a topological
phase. I must emphasize that Berry's phase and its generalization,
the geometric or Aharonov-Anandan phase are by no means
topological. By this I mean that unlike topological phases such as
the Aharonov-Bohm phase, the geometric phase does depend on
the specific shape of the curve in the parameter space or the
projective Hilbert space. The qualification {\em geometric} means
that the geometric phase is independent of the parameterization
of these curves.  In the case of topological phases, however, they
are also invariant under arbitrary continuous deformations of these
curves, alternatively the  Hamiltonian. The relation between the
geometric and the topological phases
is that the latter is a special case of the former.

In view of these considerations, Wu's article~\cite{wu} seems to
lack a logical foundation. Specifically, his claim made in
the abstract of his article, namely: ``This paper will show that, for
some specific Hamiltonians, cyclic evolution may occur, even if
the initial wave function is not one of the eigenfunctions of the  
evolution
operator,'' and repeated in the Summary (section VI): ``This paper  
has
shown that, for some special Hamiltonians, cyclic evolution occurs,
even if the initial wave function is not one of the eigenfunctions of  
the
evolution operator,''  is absolutely wrong.

\end{document}